\documentclass[prl,aps,showpacs,onecolumn,floatfix]{revtex4}
\usepackage{epsfig} \usepackage{graphics} \usepackage{bm}
\usepackage{amssymb}
\usepackage{graphicx}
\begin{document}

\preprint{Nature-2019}

\title{Inequivalence between gravitational mass and energy
of a composite quantum body in general relativity}

\author{Andrei G. Lebed$^*$}

\affiliation{Department of Physics, University of Arizona, 1118 E.
4-th Street, Tucson, Arizona 85721, USA; \
lebed@physics.arizona.edu}

\begin{abstract}
We consider the so-called semiclassical variant of general
relativity, where gravitational field is not quantized but matter
is quantized, for the simplest composite quantum body - a hydrogen
atom. We create a stationary electron quantum state in the atom in
the absence of gravitational field and study its time evolution in
the presence of the field, using the local Lorentz invariance
property of spacetime. It is shown that this state with a definite
energy in the absence of gravitational field is not anymore a
stationary state in the field. Therefore, quantum measurements of
passive gravitational mass of electron in a hydrogen atom can give
the following quantized values, $m_n = m_e + E_n/c^2$, where $m_e$
is the bare electron mass and $E_n$ is its energy level in the
atom. We discuss some difficulties in the possible experimental
observations of this mass quantization phenomenon.
\end{abstract}


\maketitle


Equivalence principle (EP) between gravitational and
inertial masses in a combination with the local Lorentz
invariance of spacetime is known to be a keystone of the
classical general relativity [1,2]. In the current scientific
literature, there exists a wide discussion if it can survive in
the possible quantum theory of gravity (see, for example, the
recent papers [3-5]). Since the quantum gravitation theory has
not been elaborated yet, the EP is often
studied in framework of the so-called semiclassical approach to
quantum gravity, where gravitational field is not quantized, but
the matter is quantized (see Refs. [3-5]). Note that the EP for a
composite body is not a trivial notion even in general relativity
in the absence of quantum effects. Indeed, as shown in Ref.[6-8],
external gravitational field is coupled not directly with energy
of a composite body but with the combination $R+3K+2P$, where $R$, $K$,
and $P$ are rest, kinetic, and potential energies, respectively. 
As mentioned in Ref.[8] and considered in detail in Ref.[9], the above mentioned
combination can be changed into expected total energy, if we
choose the proper local coordinates, where the interval has the
Minkowski's form. Therefore, in classical general relativity passive
gravitational mass is equivalent to inertial one for a composite body
[8,9], as expected. On the other hand, as
shown in Ref.[7], active gravitational mass of a composite
classical body is equivalent to its energy only after averaging
the gravitational mass over time. Semiclassical analysis [5] of
the Einstein's field equation has shown that the expectation
values of active gravitational mass and energy are equivalent only
for stationary quantum states of a composite quantum body.
Situation is different for coherent quantum superpositions of
stationary quantum states, where the expectation values of active
gravitational mass can oscillate in time even for superpositions
with constant expectation values of energy. The results [5] are
against the equivalence of active gravitational mass and energy
even at macroscopic level in quantum gravity, which has to modify
the EP.

The goal of our paper is to show theoretically that in
semiclassical gravity passive gravitational mass of a composite
quantum body is not equivalent to its energy at least at a
microscopic level. We start from quantum state with a definite electron
energy, $E_1$, in the absence of external gravitational field and show
that a measurement of the mass in the field can give a quantized
value, different from expected one, $m_n \neq m_e+\frac{E_1}{c^2}$,
although the corresponding probability is small. During our
calculations we use property of the local Lorentz invariance of a
spacetime in general relativity as well as consider passive
gravitational mass to be a quantity proportional to weight of a
composite body whose center of mass is fixed in gravitational
field by some forces of non-gravitational origin.

Suppose that at $t<0$ there is no gravitational field and electron
is in ground state of a hydrogen atom, characterizing by the
following wave function:
\begin{equation}
\Psi_1(r,t) = \exp \biggl( \frac{-i m_e c^2 t}{ \hbar} \biggl)
\exp \biggl(\frac{-i E_1 t}{ \hbar} \biggl) \Psi_1(r),
\end{equation}
which is solution of the corresponding Schr\"{o}dinger equation:
\begin{equation}
i \hbar \frac{\partial \Psi_1(r,t)}{\partial t} = \biggl[ m_e c^2
- \frac{\hbar^2}{2m_e} \biggl( \frac{\partial^2}{ \partial x^2} +
\frac{\partial^2}{ \partial y^2} + \frac{\partial^2}{ \partial
z^2} \biggl) - \frac{e^2}{r} \biggl] \Psi_1(r,t).
\end{equation}
[Here $E_1$ is electron ground state energy, $r$ is a distance
between electron with coordinates $(x,y,z)$ and proton; $\hbar$ is
the Planck constant, $c$ is the velocity of light.]

At $t=0$, let us switch on a weak centrosymmetric  (e.g. the
Earth's) gravitational field, fixing a position of center of mass
of the atom (i.e., proton) in the field by some non-gravitational forces. It
is known that in a weak field approximation curved spacetime is
characterized by the following interval [1,2]:
\begin{equation}
d s^2 = -\biggl(1 + 2 \frac{\phi}{c^2} \biggl)(cdt)^2 + \biggl(1 -
2 \frac{\phi}{c^2} \biggl) (dx^2 +dy^2+dz^2 ), \ \ \phi = -
\frac{GM}{R} .
\end{equation}
Below we introduce the proper local coordinates,
\begin{equation}
\tilde t = \biggl( 1 + \frac{\phi}{c^2} \biggl)t, \ \ \tilde x =
\biggl( 1 - \frac{\phi}{c^2} \biggl)x, \ \ \tilde y = \biggl( 1 -
\frac{\phi}{c^2} \biggl) y, \ \ \tilde z = \biggl( 1
-\frac{\phi}{c^2} \biggl) z ,
\end{equation}
where interval has the Minkowski's form [1,2],
\begin{equation}
d \tilde s^2 = -(c d \tilde t)^2 + (d \tilde x^2 +d \tilde y^2 + d
\tilde z^2).
\end{equation}
[Here, we stress that, since we are interested in calculating some
quantum transition amplitudes with the first order accuracy with
respect to the small parameter, $|\frac{\phi}{c^2}| \ll 1$, we
disregard in Eqs.(3)-(5) and therein below all terms of the order
of $\frac{\phi^2}{c^4}$. We pay attention that near the Earth's
surface the above discussed parameter is small and equal to
$|\frac{\phi}{c^2}| \sim 10^{-9}$.]

Due to the local Lorentz invariance of a spacetime in general
relativity, if we disregard the so-called tidal terms in the
Hamiltonian [i.e., if we don't differentiate the potential
$\phi(R)$], then new wave functions, written in the local proper
coordinates (4) (with fixed proton's position), satisfy at
$t,\tilde t >0$ the similar Schr\"{o}dinger equation:
\begin{equation}
i \hbar \frac{\partial \tilde \Psi(\tilde r,\tilde t)}{\partial
\tilde t} = \biggl[ m_e c^2 - \frac{\hbar^2}{2m_e} \biggl(
\frac{\partial^2}{
\partial \tilde x^2} + \frac{\partial^2}{ \partial \tilde y^2} +
\frac{\partial^2}{\partial \tilde z^2} \biggl) - \frac{e^2}{\tilde
r} \biggl] \tilde \Psi(\tilde r,\tilde t).
\end{equation}
[Note that it is easy to show that the above disregarded tidal
terms have relative order of $\frac{r_0}{R_0}$, where $r_0$ is the
Bohr radius and $R_0$ is distance between a hydrogen atom and a
center of source of gravitational field. Near the Earth's surface
they are very small and are of the relative order of
$\frac{r_0}{R_0} \sim 10^{-17}$.] $\\$
We stress that it is very
important that the wave function (1) is not a solution of the
Schr\"{o}dinger equation (6) anymore and, thus, is not
characterized by definite energy and weight in the gravitational
field (3). Moreover, a general solution of Eq.(6) can be written
in the proper local coordinates in the following way:
\begin{equation}
\tilde \Psi(\tilde r, \tilde t) = \exp \biggl(\frac{-im_e c^2
\tilde t}{\hbar}\biggl) \sum^{\infty}_{n = 1} \tilde a_n
\Psi_n(\tilde r) \exp\biggl(\frac{-i E_n \tilde t}{\hbar}\biggl),
\end{equation}
where the wave functions $\Psi_n(\tilde r)$ are solutions for the
so-called $nS$ atomic orbitals of a hydrogen atom with energies
$E_n$ [10] and are normalized in the proper local space,
\begin{equation}
\int  \Psi^2_n(\tilde r) \ d^3 \tilde r = 1.
\end{equation}
[Here we stress that, as possible to show, only $1S \rightarrow
nS$ quantum transitions amplitudes are non-zero in a hydrogen atom
in the gravitational field (3), which correspond only to real wave
functions. Therefore, we keep in Eq.(7) only $nS$ atomic orbitals
and everywhere below disregard difference between $\Psi_n(r)$ and
$\Psi^*_n(r) = \Psi_n(r)$.] $\\$ Note that the normalized wave function (1) can
be rewritten in the proper local spacetime coordinates (4) in the
following way:
\begin{equation}
\Psi_1(\tilde r,\tilde t) = \exp \biggl[\frac{-im_e c^2
(1-\frac{\phi}{c^2}) \tilde t}{\hbar}\biggl] \exp \biggl[\frac{-i
E_1 (1-\frac{\phi}{c^2}) \tilde t}{\hbar} \biggl]
\biggl(1+\frac{\phi}{c^2}\biggl)^{3/2} \Psi_1
\biggl[\biggl(1+\frac{\phi}{c^2} \biggl) \tilde r \biggl]
 \ ,
\end{equation}

It is important that the gravitational field (3) can be considered
as a sudden perturbation to the Hamiltonian (2), therefore, at
$t=\tilde t = 0$ the wave functions (7) and (9) have to be equal:
\begin{equation}
\biggl( 1+\frac{\phi}{c^2} \biggl)^{3/2} \Psi_1\biggl[
\biggl( 1+\frac{\phi}{c^2} \biggl) \tilde r \biggl] =
\sum^{\infty}_{n=1} \tilde a_n \Psi_n (\tilde r).
\end{equation}
From Eq.(10), it directly follows that
\begin{equation}
\tilde a_1 = \biggl(1+\frac{\phi}{c^2}\biggl)^{3/2} \int^{\infty}_0
\Psi_1 \biggl[\biggl(1+\frac{\phi}{c^2}\biggl) \tilde r\biggl] \Psi_1(\tilde r)
\ d^3 \tilde r
\end{equation}
and
\begin{equation}
\tilde a_n = \biggl(1+\frac{\phi}{c^2}\biggl)^{3/2}
\int^{\infty}_0 \Psi_1 \biggl[\biggl(1+\frac{\phi}{c^2}\biggl)
\tilde r\biggl] \Psi_n(\tilde r) \ d^3 \tilde r , \ \ n>1.
\end{equation}
This alow us to calculate quantum
mechanical amplitudes (11) and (12) in a linear with respect to
the gravitational potential approximation,
\begin{equation}
\tilde a_1 = 1 + O \biggl( \frac{\phi^2}{c^4} \biggl),
\end{equation}
and
\begin{equation}
\tilde a_n = \biggl(\frac{\phi}{c^2} \biggl) \int^{\infty}_0
\biggl[\frac{d \Psi_1(\tilde r)}{d \tilde r} \biggl] \tilde r \Psi_n(\tilde r)
d^3 \tilde r, \ \ n>1.
\end{equation}
We stress that the wave function (7) is a series of wave functions,
which have definite weights in the gravitational field (3). This
means that they are characterized by definite passive
gravitational masses,
\begin{equation}
m_n = m_e + \frac{E_n}{c^2}.
\end{equation}
In accordance with the most general properties of quantum
mechanics, this means that, if we do a measurement of
gravitational mass for wave function (1) and (9), we obtain
quantum values (15) with the probabilities: $\tilde P_n = |\tilde
a_n|^2$, where $\tilde a_n$ are given by Eqs.(13) and (14). 

Let us show that
\begin{equation}
\int^{\infty}_0 \biggl[\frac{d \Psi_1(\tilde r)}{d \tilde r} \biggl] \tilde r \Psi_n(\tilde r)
d^3 \tilde r = \frac{V_{1n}}{E_n-E_1}, \ \ n>1,
\end{equation}
where $\hat V(\tilde r)$ is the so-called virial operator [10]:
\begin{equation}
\hat V(r) = - 2 \frac{\hbar^2}{2m_e} \biggl( \frac{\partial^2}{
\partial \tilde x^2} + \frac{\partial^2}{ \partial \tilde y^2} +
\frac{\partial^2}{ \partial \tilde z^2} \biggl) - \frac{e^2}{\tilde r},
\end{equation}
and
\begin{equation}
V_{1,n}=\int^{\infty}_0 \Psi_1(\tilde r) \hat V(\tilde r) \Psi_n(\tilde r)
d^3 \tilde r.
\end{equation}
To this end, we rewrite the Schr\"{o}dinger equation in
gravitational field (6) in terms of the initial coordinates
$(x,y,z)$:
\begin{equation}
(m_e c^2 + E_1)
\Psi_1\biggl[\biggl(1-\frac{\phi}{c^2}\biggl)r\biggl] = \biggl[
m_e c^2 - \frac{1}{(1-\phi/c^2)^2} \frac{\hbar^2}{2m} \biggl(
\frac{\partial^2}{
\partial x^2} + \frac{\partial^2}{ \partial y^2} +
\frac{\partial^2}{\partial z^2} \biggl) - \frac{1}{(1-\phi/c^2)}
\frac{e^2}{ r} \biggl]
\Psi_1\biggl[\biggl(1-\frac{\phi}{c^2}\biggl)r\biggl] .
\end{equation}
Then, keeping as usual only terms of the first order with respect
to the small parameter $|\frac{\phi}{c^2}| \ll 1$, we obtain:
\begin{equation}
E_1 \Psi_1(r) - \frac{\phi}{c^2} E_1 r \biggl[\frac{d \Psi_1(r)}{dr} \biggl]=
\biggl[ - \frac{\hbar^2}{2m_e} \biggl( \frac{\partial^2}{
\partial x^2} + \frac{\partial^2}{ \partial y^2} +
\frac{\partial^2}{\partial z^2} \biggl) - \frac{e^2}{r} + \frac{\phi}{c^2} \hat
V(r) \biggl] \biggl[ \Psi_1(r) - \frac{\phi}{c^2} r \biggl[\frac{d
\Psi_1(r)}{dr} \biggl]\biggl],
\end{equation}
and as a result
\begin{equation}
- E_1 r \biggl[\frac{d \Psi_1(r)}{dr} \biggl] = \biggl[ -
\frac{\hbar^2}{2m_e} \biggl( \frac{\partial^2}{
\partial x^2} + \frac{\partial^2}{ \partial y^2} +
\frac{\partial^2}{\partial z^2} \biggl) - \frac{e^2}{r} \biggl]
\biggl[- r \frac{d \Psi_1(r)}{dr} \biggl] +\hat V(r) \Psi_1 (r).
\end{equation}
Let us multiply Eq.(21) on $\Psi_1(r)$ and integrate over space,
\begin{equation}
- E_1 \int^{\infty}_{0} \Psi_n(r) r \biggl[\frac{d \Psi_1(r)}{dr}
\biggl] d^3r = \int^{\infty}_{0} \Psi_n(r) \biggl[ -
\frac{\hbar^2}{2m_e} \biggl( \frac{\partial^2}{
\partial x^2} + \frac{\partial^2}{ \partial y^2} +
\frac{\partial^2}{\partial z^2} \biggl) - \frac{e^2}{r} \biggl]
\biggl[- r \frac{d \Psi_1(r)}{dr} \biggl]d^3r +\int^{\infty}_{0}
\Psi_n(r)\hat V(r) \Psi_1 (r)d^3r.
\end{equation}
Taking into account that the Hamiltonian operator in the absence
of gravitational field (2) is the Hermitian one, we rewrite
Eq.(22) as
\begin{equation}
E_1 \int^{\infty}_{0} \Psi_n(r) r \biggl[\frac{d \Psi_1(r)}{dr}
\biggl] d^3r = E_n \int^{\infty}_{0} \Psi_n(r) r \biggl[\frac{d
\Psi_1(r)}{dr} \biggl] d^3r - \int^{\infty}_{0} \Psi_1(r)\hat V(r)
\Psi_n(r)d^3r.
\end{equation}
Then, Eqs.(16)-(18) directly follow from Eq.(23).

Therefore, the calculated amplitudes (14) and the corresponding
probabilities with $n \neq 1$ can be rewritten as functions of
matrix elements (18) of the virial operator (17),
\begin{equation}
\tilde a_n = \biggl( \frac{\phi}{c^2} \biggl) \frac{V_{1,n}}{E_n-E_1}
\end{equation}
and
\begin{equation}
\tilde P_n = |\tilde a_n|^2 = \biggl( \frac{\phi}{c^2} \biggl)^2 \biggl(
\frac{V_{1,n}}{E_n-E_1} \biggl)^2.
\end{equation}
Note that near the Earth's surface, where $\frac{\phi^2}{c^4}
\approx 0.49 \times 10^{-18}$, the probability for $n=2$ in a
hydrogen atom can be calculated as
\begin{equation}
\tilde P_2 = |\tilde a_2|^2 = 1.5 \times 10^{-19},
\end{equation}
where
\begin{equation}
 \frac{V_{1,2}}{E_2-E_1} = 0.56.
\end{equation}
It is important that non-zero matrix elements (18) of the virial 
operator (17) for $n \neq 1$ are also responsible for breakdown
of the equivalence between active gravitational mass and energy for
a quantum body with internal degrees of freedom [5].

To summarize, we have demonstrated that passive gravitational mass
and energy are not equivalent in general relativity, if we take
into account basic quantum effects. The suggested phenomenon is
due to squeezing a space by gravitational field and is not
appropriate for inertial mass. Therefore, in fact we have
established inequality of gravitational and inertial masses. We do
not intend to consider some concrete experiments in the framework
of the Letter. We just discuss below several issues of experimental
interest. First of all, we point out that all energy levels with
$n>1$ are quasi-stationary and decay in time. Therefore, all mass
quantization events can be observed by optical methods, which are
currently very sensitive. Moreover, the probability (26) is not
too small since the so-called Avogadro number is $N_A \approx
6\times 10^{23}$ and, thus, we may have a large number of the
emitted photons, if we take a macroscopic number of the atoms. In
our opinion, the main experimental difficulty is that the change
of the gravitational potential has to be quick enough to generate
transitions between states in a quantum system. In our case, it
has to be of the order of $\delta t \sim \frac{2\pi \hbar}{E_1 -
E_2} \sim 10^{-15} s$. Of course, there exist quantum systems,
where the corresponding times are much longer and which, thus, are
more perspective for experimental studies. In conclusion, we
stress that, in the Letter, we have used non-relativistic
Schr\"{o}dinger equation, since we are interested in couplings of
non-relativistic electron kinetic and potential energies with
external gravitational field. We recall that relativistic
corrections are small in a hydrogen atom [11] and can just
slightly change our final results (24)-(27).

We are thankful to N.N. Bagmet, V.A. Belinski, Steven Carlip,
Fulvio Melia, Douglas Singleton, and V.E. Zakharov for fruitful
and useful discussions.

$^*$Also at: L.D. Landau Institute for Theoretical Physics, RAS,
2 Kosygina Street, Moscow 117334, Russia.



\vspace{6pt}



\section{References}

\begin{thebibliography}{6}


\bibitem{Misner}
C.W. Misner, K.S. Thorne, and J.A. Wheeler, {\it Gravitation}
(W.H. Freeman and Co, New York, USA, 1973).

\bibitem{Landau-1}
L.D. Landau and E.M. Lifshitz, {\it The Classical Theory of
Fields} (Butterworth-Heinemann, Oxford, UK, 2003).

\bibitem{Zych-1}
M. Zych and C. Brukner, Nature Phys. \textbf{14}, 1027 (2018).

\bibitem{Zych-2}
G. Rosi, D'Amico, L.Cacciapuoti, F. Sorrentino, M.Prevedelli, M.
Zych, C. Brukner, G.M. Tito, Nature Commun. \textbf{8}, 15529
(2016).

\bibitem{Lebed}
A.G. Lebed, J. Phys.: Conf. Ser. \textbf{738}, 012036 (2016).


\bibitem{Misner} C.W. Misner and P. Putnam, {\it Phys. Rev.} {\bf 116}, 1045 (1959).


\bibitem{Nord}
K. Nordtvedt, {\it Class. Quantum Grav.} \textbf{11}, A119 (1994).

\bibitem{Carlip}
S. Carlip, {\it Am. J. Phys.} \textbf{66}, 409 (1998).

\bibitem{Zych-3}
M. Zyck, L. Rudnicki, and I. Pikovski, arXiv:1808.05831v1 (2018).

\bibitem{Park}
D. Park, {\it Introduction to the Quantum Theory}, 3rd edn. (Dover
Publications, New York, USA, 2005).



\bibitem{Schwabl}
F. Schwabl, {\it Advanced Quantum Mechanics}
(Springer-Verlag, Berlin, Germany, 2005).




\end{thebibliography}
\end{document}